\documentclass[12pt]{iopart}
\usepackage{graphicx}
\usepackage{cite}

\begin{document}

\title{Wideband trapping of light by edge states in honeycomb photonic crystals}
\author{Chunfang Ouyang$^{1}$, Dezhuan Han$^{1}$, Fangyuan Zhao$^{2}$,
Xinhua Hu$^{2}$, Xiaohan Liu$^{1}$ and Jian Zi$^{1}$}
\address{
$^{1}$Department of Physics, Key Laboratory of Micro \& Nano Photonic Structures
(Ministry of Education), and Key Laboratory of Surface Physics, Fudan
University, Shanghai 200433, P. R. China \\
$^{2}$Department of Materials Science and Laboratory of Advanced Materials,
Fudan University, Shanghai 200433, P. R. China}
\eads{\mailto{huxh@fudan.edu.cn} and \mailto{jzi@fudan.edu.cn}}

\date{\today }

\begin{abstract}
We study theoretically light propagations at the zigzag edge of a honeycomb
photonic crystal consisting of dielectric rods in air, analogous to graphene.
Within the photonic band gap of the honeycomb photonic crystal, a unimodal
edge state may exist with a sharp confinement of optical fields. Its dispersion
can be tuned simply by adjusting the radius of the edge rods.
For the edge rods with a graded variation in radius along the edge direction,
we show numerically that light beams of different frequencies can be trapped
sharply in different spatial locations, rendering wideband trapping of light.
\end{abstract}

\submitto{\JPCM}
\maketitle

Graphene, a flat monolayer of graphite with carbon atoms arranged in a honeycomb
lattice, has many interesting physical properties with great promise in nanoelectronics
owing to its unique electronic properties, e.g., conical dispersion (Dirac cones)
in the electronic band structure \cite{cas:09}. Inspired by graphene, its photonic analog,
namely, two-dimensional photonic crystals \cite{yab:87,joh:87,joa:08} that
possess Dirac cones in photonic band structures, have drawn considerable
attention in recent years \cite{pel:07,sep:07,hal:08,zha:08,han:09,ben:09,zan:10,hua:11}.
The existence of Dirac cones may lead to many
interesting optical phenomena such as conical diffraction and gap solitons \cite{pel:07},
pseudodiffusive transport of light \cite{sep:07,zan:10}, zitterbewegung of photons
\cite{zha:08} and the realization of zero-refractive-index materials \cite{hua:11}.

Wideband trapping of light, also called ``Rainbow'' trapping, is an optical
phenomenon that light of different frequencies can be slowed
down and stopped at different spatial positions in specially designed
waveguides \cite{tsa:07}. It has many potential applications
ranging from optical information storage to enhanced light-matter interactions
\cite{vla:05}. Several structures for realizing wideband trapping have been proposed,
including metamaterial waveguides \cite{tsa:07,zha:09}, plasmonic waveguides
\cite{sto:04}, metal-coated optical nano waveguides
\cite{smo:10} and plasmonic gratings \cite{gan:08,gan:11}. However, these
structures may suffer inevitably a serious problem of optical loss due to
constituting metals, which are rather lossy especially in the
visible regime. In contrast, dielectric photonic crystal waveguides can also
slow light with a very small loss \cite{ger:05,kra:07}, which could be exploited
in light trapping \cite{mor:05}. But they are usually multimodal \cite{kra:07},
which may lead to undesirable mode coupling.

By breaking the inversion symmetry of photonic crystals that possess Dirac cones,
a photonic band gap (PBG) around Dirac points may open up \cite{cho:08}.
With the presence of edges, edge states within the PBG may appear with a
sharp confinement of optical fields at the edges \cite{ouy:11}.
With further breaking the time-reversal symmetry, confined edge states
become even unidirectional, showing robustness against scattering or disorder
\cite{hal:08,wan:08,och:10,fu:10,poo:11}. In this paper, we show theoretically
that edge states in honeycomb photonic crystals (HPCs) can be exploited in
realizing wideband trapping of light. Since no metal is introduced, the optical
loss should be much lower than those in metamaterial and plasmonic waveguides.
Compared with photonic crystal waveguides, our system is unimodal and
more compact as well. Although exemplified by HPCs with rods being infinitely
long, our idea can be applied to two-dimensional photonic crystals with finite
rod lengths or membrane photonic crystals with drilled holes \cite{ger:05,kra:07,mor:05}.

\begin{figure}[tbp]
\centerline{\includegraphics[angle=0,width=10cm]{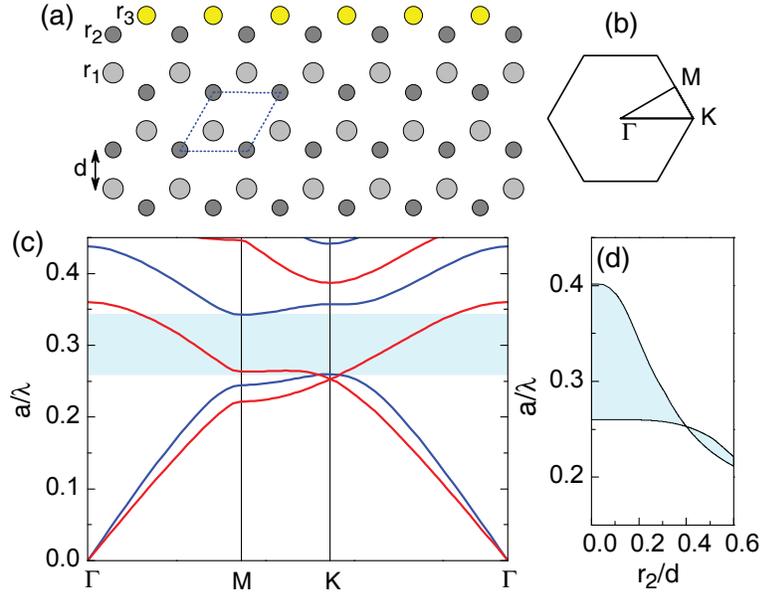}}
\caption{(a) Top view of an HPC with a zigzag edge. Light and dark gray solid
circles represent different sets of rods in a honeycomb lattice. Edge rods are
marked by yellow solid circles. Doted lines represent the unit cell of the HPC.
(b) First Brillouin zone of the honeycomb lattice with $\Gamma$, $M$ and $K$
representing points of high symmetry. (c) Photonic band structures of HPCs.
Red lines are those for $r_{1}=r_{2}=0.4d$, while blue lines for $r_{1}=0.4d$ and
$r_{2}=0.2d$. The colored area stands for the PBG of the latter HPC.
(d) PBG (colored area) map as a function of $r_{2}$ for fixed $r_{1}=0.4d$.
Frequency is in units of $a/\lambda$, where $a$ is the lattice constant and
$\lambda$ is the wavelength in vacuum.}
\label{fig1}
\end{figure}

HPCs under study consist of dielectric rods which are arranged in
a honeycomb lattice in air, analogous to graphene, as schematically shown in
figure \ref{fig1}(a). The inter-distance between adjacent rods is denoted by $d$.
In an infinite HPC, there are two different sets of rods whose radii are denoted
respectively by $r_{1}$ and $r_{2}$. The radius of edge rods, which are changeable,
is denoted by $r_{3}$.
The lattice constant of an HPC is thus $a=\sqrt{3}d$. Without loss of generality, the
dielectric constant of all rods is taken to be 11.4, a typical value for
silicon at telecommunication wavelengths. In our following discussions,
we only consider the transverse magnetic polarization, i.e., the case that the electric
field is parallel to the rods.

In figure \ref{fig1}(c), photonic band structures of infinite HPCs,
calculated by a plane-wave expansion method \cite{joa:08}, are
shown. For an HPC possessing the inversion symmetry, i.e.,
$r_{1}=r_{2}$, Dirac points exist where the first and second photonic bands
cross each other at the $K$ point with a reduced frequency of $a/\lambda=0.253$.
With broken inversion symmetry, i.e., $r_{1}\neq r_{2}$, the degeneracy at the
$K$ point will be lifted and a PBG will open up \cite{cho:08}. The lower
(upper) edge of the PBG is defined by the first (second) photonic band at the $K$
($M$) point.

To attain an HPC with an optimal or desired PBG, one can tune the radius of
the rods or adjust the dielectric constant of the rods. In figure \ref{fig1}(d),
the PBG map as a function of $r_{2}$ for fixed $r_{1}=0.4d$ is shown.
For $r_{1}=r_{2}=0.4d$, there is no PBG due to the existence of the Dirac points.
For $r_{1}\neq r_{2}$, a PBG always exists. For $r_{1}=0.4d$
and $r_{2}=0.2d$, the PBG ranges in reduced frequency $a/\lambda$ from
0.26 to 0.343. For $r_{1}=0.4d$ and $r_{2}=0$, the relative PBG width,
$\Delta \omega/\omega_{\rm m}$, can be as high as $43\%$,
where $\Delta \omega$ is the PBG width and $\omega_{\rm m}$ is the mid-gap
frequency.

\begin{figure}[tbp]
\centerline{\includegraphics[angle=0,width=10cm]{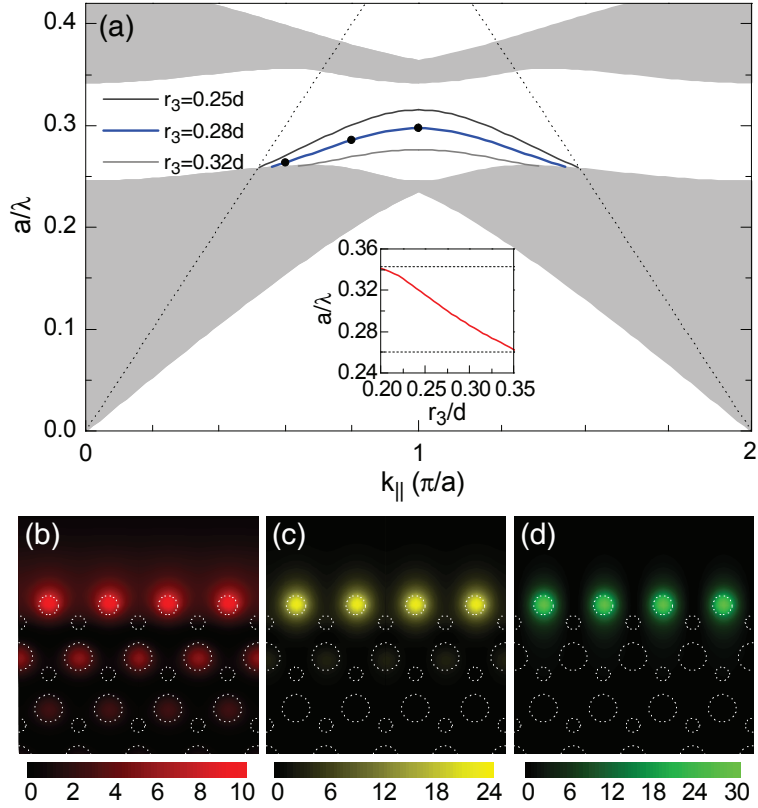}}
\caption{(a) Calculated dispersion of edge states for an HPC with a zigzag edge
as shown in figure \ref{fig1}(a), where $k_{||}$ is the wavevector along the edge
direction in units of $\pi/a$. The radius parameters for the HPC are $r_{1}=0.4d$
and $r_{2}=0.2d$. Gray areas represent projected photonic bands. Gray, blue,
and dark gray lines represent the edge states for the edge rods with different
radius. Dotted straight lines stand for the light line in vacuum. Inset shows the
cutoff frequency (at $k_{||}=\pi/a$) as a function of $r_{3}$. Two doted lines
defines the range of the PBG. (b)--(d) Distributions of the electric field
intensity $|E|^{2}$ for the edge state with $r_{3}=0.28d$ at (b) $k_{||}=0.6\pi/a$,
(c) $0.8\pi/a$, and (d) $\pi/a$, marked by dots in (a). Doted circles stand for
 the rods.}
\label{fig2}
\end{figure}

Figure \ref{fig2}(a) shows the dispersion of edge states of an HPC with a zigzag
edge, calculated by the plane-wave expansion method \cite{joa:08}. In our
calculations, a supercell with a size of $a$ along the edge direction and $46d$ along
the direction perpendicular to the edge direction is used. Periodical boundary
conditions are imposed in both directions. The supercell contains a
finite HPC with a thickness of $35d$ along the direction perpendicular to the edge
direction, corresponding to 48 rows of rods including the edge rods.
Since the separation of the two zigzag edges are large enough, there is no
coupling between the two edges. The resulting dispersion
can thus be considered as that for a single zigzag edge of a semi-infinite HPC.
Projected photonic bands are obtained by projecting the bulk photonic
band structure over the wavevector along the edge
direction, $k_{||}$. In addition to photonic bands, for a given radius
of the edge rods $r_{3}$, an edge state may appear between the first and second
photonic bands. Interestingly, this edge state is unimodal, whose dispersion
depends on $r_{3}$.

Even for different values of $r_{3}$, the dispersion of the edge state has a
similar behaviour: its frequency increases with increasing $k_{||}$ up to the
Brillouin zone boundary $k_{||}=\pi/a$. At $k_{||}=\pi/a$, it takes a zero slope.
In other words, the group velocity decreases with increasing $k_{||}$ and
approaches zero at the Brillouin zone boundary. This interesting feature
could be exploited to slow down and even stop light \cite{vla:05,ger:05}.
The frequency of an edge state at $k_{||}=\pi/a$ is dubbed cutoff frequency
since the frequencies of the edge state at other wavevectors are always
smaller than the cutoff frequency. For different values of $r_{3}$, the
corresponding cutoff frequency is different, showing a monotonic decrease
with increasing $r_{3}$, as shown in the inset of figure \ref{fig2}(a).

Note that edge states lie below the light line, implying that they are
evanescent waves. In Figs. \ref{fig2}(b)--(d), the field distributions
for the edge state with $r_{3}=0.28d$ at different $k_{||}$ are shown. Obviously,
optical fields are confined to the edge rods. For the edge mode
at $k_{||}=0.6\pi/a$, it is near the first photonic band, leading to some bulk
excitations in the vicinity of the edge, which can be clearly seen from
figure \ref{fig2}(b). In contrast, for the edge modes away from the photonic bands
there are nearly no bulk excitations. As shown by us \cite{ouy:11}, these
edge modes are of subwavelength, similar to surface plasmons bounded
at metallic surfaces \cite{bar:03}.

As aforementioned, an edge state at its cutoff frequency has a zero group
velocity and the cutoff frequency decreases monotonically with increasing
$r_{3}$. These features can render wideband trapping of light by taking
advantage of edge states, provided that the edge rods are designed to
have a spatially graded variation in radius. If the grade
is small enough, the graded edge can be considered approximately as
a series of small edges, in each of which the edge rods possess
a constant radius. As a result, each location seems to have a different
cutoff frequency. For a wave propagating along such a graded edge, its group velocity
will be gradually reduced along the edge direction and eventually approach
zero in a location where the corresponding cutoff frequency coincide with the
frequency of the wave. In other words, the wave will stop around this location.

\begin{figure}[tbp]
\centerline{\includegraphics[angle=0,width=13.5cm]{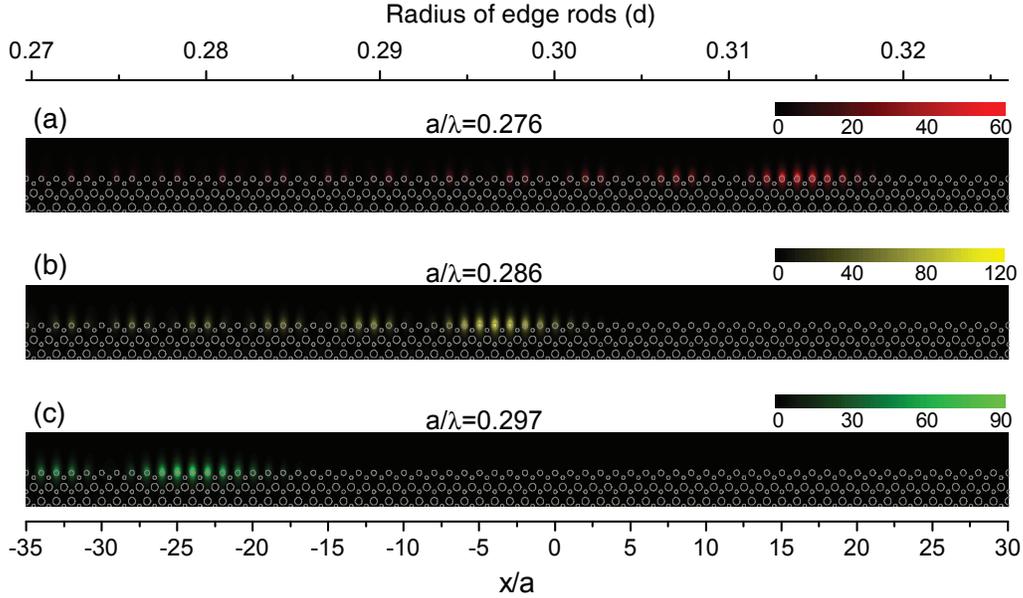}}
\caption{FDTD simulations of the electric field intensity $|E|^{2}$ for an HPC with a
graded zigzag edge at different reduced frequencies (a) 0.276, (b) 0.286, and (c)
0.297. The radius parameters for the HPC are $r_{1}=0.4d$ and $r_{2}=0.2d$.
A right-propagating Gaussian beam
of different frequencies is launched at $x=-46a$. Circles denote dielectric rods.
The bottom (top) horizontal axis indicates the position (radius of edge rods)
along the edge direction.}
\label{fig3}
\end{figure}

Figure \ref{fig3} shows light trapping for different frequencies at the graded
zigzag edge of an HPC, simulated by a finite-difference time-domain (FDTD)
method \cite{taf:95}. The radii of the edge rods are assumed to vary linearly
with the position of the edge rods, namely,
$r_{3}=0.3d+5\times10^{-4}x$, where $x$ is the position along the edge direction.
In our FDTD simulations, a continuous Gaussian beam with an intensity of 1 and
a width of $2d$ is launched at $x=-46a$. Perfectly matched layer
absorbing boundaries \cite{ber:94} are used.

Obviously, from figure \ref{fig3} light beams of different frequency can be trapped
spatially in different locations. The scenario is as follows. For a
right-propagating Gaussian beam, it will propagate rightwards after
launching. The beam will be slowed down towards right since its group velocity
decreases rightwards due to the reduction of the radius of the
edge rods. It will eventually be stopped around the location where the corresponding
cutoff frequency of the location coincides with the beam frequency.
By inspecting the beam with a reduced frequency of 0.286, for example, the maximum
intensity of the standstill beam occurs around the position $x\approx-4a$. The radius of
the edge rods around this location is about $0.297d$, very close to but a bit smaller
than the radius of $0.3d$ that offers a cutoff frequency of 0.286 [see the inset of figure
\ref{fig2}(a)].
For the standstill beam, there are some small additional peaks on the
left side of the main peak. This is because the right-propagating wave may get
reflected more or less since the grade of the edge rods is not small enough.
The reflected and the right-propagating waves may interfere with each other,
leading to these small additional peaks, which could be suppressed in principle
by using a finer grade.

In summary, we studied theoretically the edge states at the zigzag edge of HPCs
consisting of dielectric rods in air for the transverse magnetic polarization.
With broken inversion symmetry, a PBG exists between the first and second
photonic bands. A unimodal edge state may appear within the PBG, whose
dispersion and cutoff frequency can be tuned by changing the radius of the edge
rods. For a zigzag edge with a gradually graded radius in edge rods,
light beams of different frequencies can be slowed down and trapped spatially in
different locations. Our results show that edge states in HPCs could be exploited
in optical storage and enhanced light-matter interactions.\\

This work was supported by the 973 Program (Grant nos. 2011CB922004,
2012CB921604 and 2013CB632701) and the NSFC.

\clearpage
\end{document}